\documentclass[dvips]{article}
\usepackage{supertabular,lscape,epsfig}
\usepackage{amssymb}
\usepackage{amsmath}

\DeclareSymbolFont{ppa}{OT1}{ppl}{m}{it}
\DeclareMathSymbol{\vv}{\mathalpha}{ppa}{'166}

\begin{document}

\newcommand{\dd}{\,{\rm d}}
\newcommand{\ie}{{\it i.e.},\,}
\newcommand{\etal}{{\it et al.\ }}
\newcommand{\eg}{{\it e.g.},\,}
\newcommand{\cf}{{\it cf.\ }}
\newcommand{\vs}{{\it vs.\ }}
\newcommand{\zdot}{\makebox[0pt][l]{.}}
\newcommand{\up}[1]{\ifmmode^{\rm #1}\else$^{\rm #1}$\fi}
\newcommand{\dn}[1]{\ifmmode_{\rm #1}\else$_{\rm #1}$\fi}
\newcommand{\upd}{\up{d}}
\newcommand{\uph}{\up{h}}
\newcommand{\upm}{\up{m}}
\newcommand{\ups}{\up{s}}
\newcommand{\arcd}{\ifmmode^{\circ}\else$^{\circ}$\fi}
\newcommand{\arcm}{\ifmmode{'}\else$'$\fi}
\newcommand{\arcs}{\ifmmode{''}\else$''$\fi}
\newcommand{\MS}{{\rm M}\ifmmode_{\odot}\else$_{\odot}$\fi}
\newcommand{\RS}{{\rm R}\ifmmode_{\odot}\else$_{\odot}$\fi}
\newcommand{\LS}{{\rm L}\ifmmode_{\odot}\else$_{\odot}$\fi}

\newcommand{\Abstract}[2]{{\footnotesize\begin{center}ABSTRACT\end{center}
\vspace{1mm}\par#1\par
\noindent
{~}{\it #2}}}

\newcommand{\TabCap}[2]{\begin{center}\parbox[t]{#1}{\begin{center}
  \small {\spaceskip 2pt plus 1pt minus 1pt T a b l e}
  \refstepcounter{table}\thetable \\[2mm]
  \footnotesize #2 \end{center}}\end{center}}

\newcommand{\TableSep}[2]{\begin{table}[p]\vspace{#1}
\TabCap{#2}\end{table}}

\newcommand{\FigCap}[1]{\footnotesize\par\noindent Fig.\  %
  \refstepcounter{figure}\thefigure. #1\par}

\newcommand{\TableFont}{\footnotesize}
\newcommand{\TableFontIt}{\ttit}
\newcommand{\SetTableFont}[1]{\renewcommand{\TableFont}{#1}}

\newcommand{\MakeTable}[4]{\begin{table}[htb]\TabCap{#2}{#3}
  \begin{center} \TableFont \begin{tabular}{#1} #4 
  \end{tabular}\end{center}\end{table}}

\newcommand{\MakeTableSep}[4]{\begin{table}[p]\TabCap{#2}{#3}
  \begin{center} \TableFont \begin{tabular}{#1} #4 
  \end{tabular}\end{center}\end{table}}

\newenvironment{references}%
{
\footnotesize \frenchspacing
\renewcommand{\thesection}{}
\renewcommand{\in}{{\rm in }}
\renewcommand{\AA}{Astron.\ Astrophys.}
\newcommand{\AAS}{Astron.~Astrophys.~Suppl.~Ser.}
\newcommand{\ApJ}{Astrophys.\ J.}
\newcommand{\ApJS}{Astrophys.\ J.~Suppl.~Ser.}
\newcommand{\ApJL}{Astrophys.\ J.~Letters}
\newcommand{\AJ}{Astron.\ J.}
\newcommand{\IBVS}{IBVS}
\newcommand{\PASP}{P.A.S.P.}
\newcommand{\Acta}{Acta Astron.}
\newcommand{\MNRAS}{MNRAS}
\renewcommand{\and}{{\rm and }}
\section{{\rm REFERENCES}}
\sloppy \hyphenpenalty10000
\begin{list}{}{\leftmargin1cm\listparindent-1cm
\itemindent\listparindent\parsep0pt\itemsep0pt}}%
{\end{list}\vspace{2mm}}

\def\TYLDA{~}
\newlength{\DW}
\settowidth{\DW}{0}
\newcommand{\dw}{\hspace{\DW}}

\newcommand{\refitem}[5]{\item[]{#1} #2%
\def\REFARG{#3}\ifx\REFARG\TYLDA\else, {\it#3}\fi
\def\REFARG{#4}\ifx\REFARG\TYLDA\else, {\bf#4}\fi
\def\REFARG{#5}\ifx\REFARG\TYLDA\else, {#5}\fi.}

\newcommand{\Section}[1]{\section{#1}}
\newcommand{\Subsection}[1]{\subsection{#1}}
\newcommand{\Acknow}[1]{\par\vspace{5mm}{\bf Acknowledgements.} #1}
\pagestyle{myheadings}

\newfont{\bb}{ptmbi8t at 12pt}
\newcommand{\xrule}{\rule{0pt}{2.5ex}}
\newcommand{\xxrule}{\rule[-1.8ex]{0pt}{4.5ex}}
\def\thefootnote{\fnsymbol{footnote}}
\begin{center}
{\Large\bf The Optical Gravitational Lensing Experiment.\\
\vskip3pt
Multiple Cluster Candidates in the Large Magellanic Cloud\footnote{Based on observations obtained with the 1.3~m
Warsaw telescope at the Las Campanas Observatory of the Carnegie
Institution of Washington.}}

\vskip1cm
G.~~P~i~e~t~r~z~y~\'n~s~k~i$^{1,2}$~~ and~~A.~~U~d~a~l~s~k~i$^2$
\vskip3mm
$^1$ Universidad de Concepci{\'o}n, Departamento de Fisica,
Casilla 160--C, Concepci{\'o}n, Chile\\
$^2$Warsaw University Observatory, Al.~Ujazdowskie~4, 00-478~Warszawa,
Poland\\ 
e-mail: (pietrzyn,udalski)@sirius.astrouw.edu.pl 
\end{center}

\Abstract{We present 100 multiple cluster candidates selected from the OGLE 
catalog of star clusters in the Large Magellanic Cloud. Statistical analysis 
shows that the significant fraction of these objects may constitute physical 
systems. Coeval ages of 102 components of multiple objects suggest their 
common origin. 53 components have very different ages. The comparison of the 
population of multiple clusters candidates from the SMC and the LMC shows 
that: a)~distributions of sizes and ages of multiple and single clusters from 
the Magellanic Clouds are very similar, b)~the difference of sizes of 
components of a given system is small, c)~in both distributions of separation 
of multiple clusters from the LMC and SMC two peaks are seen at about 9~pc and 
15~pc, d)~both age distributions reveal peaks around 100~Myr, which may be 
connected with the last encounter of the LMC and the SMC.}

\vskip32pt
\Section{Introduction}
\vskip9pt
Previous searches revealed that the Magellanic Clouds possess numerous 
multiple cluster candidates. Bhatia and Hatzidimitriou (1988) and Bhatia \etal 
(1991) constructed the catalog of cluster pairs from the LMC. Binary clusters 
from the SMC were described by Hatzidimitriou and Bhatia (1990). Recently, 
based on the OGLE catalog of star clusters from the SMC (Pietrzy{\'n}ski \etal 
1998), the list of multiple star cluster candidates from the central parts of 
this galaxy was presented by Pietrzy{\'n}ski and Udalski (1999a). Simple 
statistical considerations show that the number of multiple cluster candidates 
is significantly larger than the number expected from the chance line-up due 
to projection. In the case of selected cluster systems from the LMC their 
physical connections were confirmed based on detailed studies (Kontizas \etal 
1993, Vallenari \etal 1998, Leon \etal 1998, Dieball and Grebel 2000 and 
references therein). 

Having important implications on the processes of formation and evolution of 
clusters, the multiple star clusters were also subject of several theoretical 
investigations. Fujimoto and Kumai (1997) pointed out that the binary clusters 
could be formed by the oblique collisions between massive gas clouds. This 
scenario led to systems of clusters with very similar ages. Another scenario of 
tidal capture in groups of clusters was proposed by Leon \etal (1999). Such a 
mechanism explains formation of systems with members having large age 
difference. 

The Optical Gravitational Lensing Experiment (OGLE) (Udalski, Kubiak and 
Szyma{\'n}ski 1997) provided in precise {\it BVI} observations of millions of 
stars from the Magellanic Clouds (Udalski \etal 1998). The OGLE data, covering 
relatively large areas in the Clouds are very well suited for searching and 
analyzing the properties of star clusters from these galaxies. In a series of 
papers we present results of these investigations. Catalogs of clusters from 
the LMC and the SMC were presented by Pietrzy{\'n}ski \etal (1999) and 
Pietrzy{\'n}ski \etal (1998), respectively. Ages for about 700 of these 
clusters were determined using the standard procedure of isochrone fitting 
(Pietrzy{\'n}ski and Udalski 1999b, 2000). Possible multiple clusters from 
the SMC were listed by Pietrzy{\'n}ski and Udalski (1999c). The lists of 
eclipsing systems and Cepheids in coincidence with star clusters in Magellanic 
Clouds are given in Pietrzy{\'n}ski and Udalski (1999 ad). 

In this contribution we present multiple cluster candidates selected from the 
catalog of star clusters in central parts of the Large Magellanic Cloud and 
compare them with the objects from the OGLE catalog of multiple clusters from 
the SMC (Pietrzy\'nski and Udalski 1999c). 

\vskip30pt
\Section{LMC Multiple Cluster Candidates}
\vskip7pt
Following the previous catalogs, clusters with projected separations smaller 
than 18~pc, assuming the distance modulus to the LMC of 18.24~mag (Udalski 
2000), were selected from the OGLE catalog of clusters from the 5.8 square 
degrees region in the LMC (Pietrzy{\'n}ski \etal 1999). Among 745 star 
clusters we detected 73, 18, 5, 1 and 3 systems consisting of 2, 3, 4, 5 and 
6 clusters, respectively. Table~1 contains their description. Cluster 
designations, equatorial coordinates and radii were taken from the catalog of 
clusters (Pietrzy{\'n}ski \etal 1999). Ages were taken from Pietrzy{\'n}ski 
and Udalski (2000). 

\renewcommand{\arraystretch}{.95}
\MakeTableSep{|c|c|c|c|c|c|}{12.5cm}{The multiple cluster candidates from the LMC}
{\hline
Name  &  $\alpha_{2000}$ & $\delta_{2000}$ & Radius
& $\log t$ & $\sigma_{\log t}$\\
OGLE-CL-& & &[\arcs]&\\
\hline\xrule
LMC0011 & $5\uph01\upm17\zdot\ups72$ & $-67\arcd18\arcm06\zdot\arcs7$ &    21 & -- &  -- \\
LMC0012 & $5\uph01\upm22\zdot\ups46$ & $-67\arcd17\arcm41\zdot\arcs2$ &    20 & 8.7 &  0.1 \\
LMC0014 & $5\uph01\upm26\zdot\ups82$ & $-67\arcd17\arcm42\zdot\arcs6$ &    12 & -- &  -- \\
\hline\xrule
LMC0043 & $5\uph03\upm38\zdot\ups64$ & $-68\arcd58\arcm44\zdot\arcs3$ &    13 & 8.5 &  0.05 \\
LMC0044 & $5\uph03\upm42\zdot\ups13$ & $-68\arcd58\arcm06\zdot\arcs3$ &    14 & 7.7 &  0.1 \\
LMC0048 & $5\uph03\upm49\zdot\ups94$ & $-68\arcd58\arcm37\zdot\arcs5$ &    10 & -- &  -- \\
\hline\xrule
LMC0053 & $5\uph04\upm19\zdot\ups30$ & $-69\arcd21\arcm23\zdot\arcs2$ &    10 & 8.8 &  0.1 \\
LMC0057 & $5\uph04\upm24\zdot\ups94$ & $-69\arcd20\arcm59\zdot\arcs7$ &    13 & 8.85 &  0.05 \\
LMC0059 & $5\uph04\upm30\zdot\ups57$ & $-69\arcd21\arcm18\zdot\arcs3$ &    20 & 8.8 &  0.1 \\
LMC0061 & $5\uph04\upm39\zdot\ups09$ & $-69\arcd20\arcm26\zdot\arcs1$ &    25 & 8.4 &  0.1 \\
\hline\xrule
LMC0063 & $5\uph04\upm44\zdot\ups89$ & $-68\arcd59\arcm03\zdot\arcs8$ &    19 & 8.4 &  0.1 \\
LMC0064 & $5\uph04\upm50\zdot\ups43$ & $-68\arcd59\arcm16\zdot\arcs2$ &    20 & 8.4 &  0.05 \\
\hline\xrule
LMC0066 & $5\uph05\upm00\zdot\ups64$ & $-68\arcd45\arcm01\zdot\arcs3$ &    14 & 8.0 &  0.05 \\
LMC0075 & $5\uph05\upm14\zdot\ups14$ & $-68\arcd44\arcm34\zdot\arcs4$ &    18 & 7.9 &  0.05 \\
LMC0077 & $5\uph05\upm18\zdot\ups53$ & $-68\arcd43\arcm33\zdot\arcs7$ &    19 & 7.9 &  0.1 \\
LMC0078 & $5\uph05\upm19\zdot\ups18$ & $-68\arcd44\arcm14\zdot\arcs7$ &    30 & 8.0 &  0.1 \\
\hline\xrule
LMC0081 & $5\uph05\upm35\zdot\ups79$ & $-68\arcd37\arcm42\zdot\arcs5$ &    35 & 8.5 &  0.05 \\
LMC0083 & $5\uph05\upm40\zdot\ups09$ & $-68\arcd38\arcm11\zdot\arcs9$ &    25 & 7.8 &  0.05 \\
\hline\xrule
LMC0089 & $5\uph05\upm55\zdot\ups36$ & $-68\arcd57\arcm04\zdot\arcs8$ &    17 & 8.0 &  0.1 \\
LMC0092 & $5\uph06\upm02\zdot\ups27$ & $-68\arcd57\arcm22\zdot\arcs2$ &    14 & 7.2 &  0.2 \\
\hline\xrule
LMC0090 & $5\uph05\upm55\zdot\ups63$ & $-68\arcd37\arcm42\zdot\arcs8$ &    27 & 7.8 &  0.1 \\
LMC0093 & $5\uph06\upm02\zdot\ups89$ & $-68\arcd37\arcm41\zdot\arcs6$ &    25 & 8.0 &  0.05 \\
\hline\xrule
LMC0103 & $5\uph06\upm24\zdot\ups14$ & $-69\arcd34\arcm06\zdot\arcs1$ &    23 & 8.7 &  0.1 \\
LMC0108 & $5\uph06\upm33\zdot\ups66$ & $-69\arcd34\arcm05\zdot\arcs4$ &    18 & -- &  -- \\
\hline\xrule
LMC0105 & $5\uph06\upm24\zdot\ups81$ & $-68\arcd22\arcm29\zdot\arcs5$ &    20 & 8.35 &  0.1 \\
LMC0107 & $5\uph06\upm33\zdot\ups57$ & $-68\arcd21\arcm47\zdot\arcs3$ &    19 & 7.3 &  0.05 \\
\hline\xrule
LMC0133 & $5\uph07\upm55\zdot\ups46$ & $-69\arcd17\arcm57\zdot\arcs3$ &    11 & 8.3 &  0.1 \\
LMC0135 & $5\uph08\upm03\zdot\ups87$ & $-69\arcd18\arcm03\zdot\arcs7$ &    10 & -- &  -- \\
\hline\xrule
LMC0140 & $5\uph08\upm34\zdot\ups99$ & $-69\arcd10\arcm36\zdot\arcs1$ &    20 & 8.85 &  0.05 \\
LMC0141 & $5\uph08\upm43\zdot\ups59$ & $-69\arcd10\arcm58\zdot\arcs6$ &    18 & 7.7,8.2 &  0.1 \\
\hline\xrule
LMC0142 & $5\uph08\upm45\zdot\ups79$ & $-68\arcd45\arcm38\zdot\arcs6$ &    63 & 7.9 &  0.05 \\
LMC0145 & $5\uph08\upm54\zdot\ups55$ & $-68\arcd45\arcm13\zdot\arcs9$ &    29 & 8.0 &  0.1 \\
\hline\xrule
LMC0149 & $5\uph09\upm12\zdot\ups95$ & $-69\arcd17\arcm00\zdot\arcs0$ &    11 & 9.05 &  0.05 \\
LMC0151 & $5\uph09\upm14\zdot\ups13$ & $-69\arcd16\arcm00\zdot\arcs9$ &     7 & -- &  -- \\
\hline\xrule
LMC0154 & $5\uph09\upm20\zdot\ups10$ & $-68\arcd50\arcm52\zdot\arcs8$ &    43 & 7.85 &  0.05 \\
LMC0155 & $5\uph09\upm24\zdot\ups99$ & $-68\arcd51\arcm47\zdot\arcs4$ &    25 & 7.6 &  0.05 \\
LMC0156 & $5\uph09\upm28\zdot\ups43$ & $-68\arcd51\arcm01\zdot\arcs5$ &    21 & 7.8 &  0.05 \\
\hline\xrule
LMC0160 & $5\uph09\upm42\zdot\ups92$ & $-68\arcd48\arcm06\zdot\arcs5$ &    18 & 8.2 &  0.1 \\
LMC0161 & $5\uph09\upm45\zdot\ups66$ & $-68\arcd47\arcm18\zdot\arcs1$ &    22 & 8.3 &  0.1 \\
\hline\xrule
LMC0169 & $5\uph10\upm07\zdot\ups06$ & $-69\arcd05\arcm15\zdot\arcs2$ &    16 & -- &  -- \\
LMC0170 & $5\uph10\upm11\zdot\ups15$ & $-69\arcd05\arcm15\zdot\arcs0$ &    12 & 8.0 &  0.1 \\
LMC0173 & $5\uph10\upm18\zdot\ups54$ & $-69\arcd04\arcm46\zdot\arcs5$ &    10 & 7.7 &  0.1 \\
\hline}
\setcounter{table}{0}
\renewcommand{\arraystretch}{1}
\MakeTableSep{|c|c|c|c|c|c|}{12.5cm}{continued}
{\hline\xrule
Name  &  $\alpha_{2000}$ & $\delta_{2000}$ & Radius
& $\log t$ & $\sigma_{\log t}$\\
OGLE-CL-& & &[\arcs]&\\
\hline\xrule
LMC0175 & $5\uph10\upm19\zdot\ups85$ & $-69\arcd31\arcm23\zdot\arcs9$ &    11 & -- &  -- \\
LMC0181 & $5\uph10\upm30\zdot\ups90$ & $-69\arcd30\arcm59\zdot\arcs8$ &    16 & -- &  -- \\
\hline\xrule
LMC0176 & $5\uph10\upm20\zdot\ups23$ & $-68\arcd52\arcm37\zdot\arcs6$ &    19 & 8.3 &  0.1 \\
LMC0179 & $5\uph10\upm29\zdot\ups73$ & $-68\arcd52\arcm21\zdot\arcs5$ &   21 & 8.35 &  0.1\\
\hline\xrule
LMC0177 & $5\uph10\upm22\zdot\ups56$ & $-68\arcd55\arcm41\zdot\arcs8$ &    34 & 8.2 &  0.05 \\
LMC0180 & $5\uph10\upm30\zdot\ups90$ & $-68\arcd56\arcm03\zdot\arcs1$ &    27 & 8.0 &  0.2 \\
\hline\xrule
LMC0209 & $5\uph12\upm00\zdot\ups99$ & $-69\arcd12\arcm04\zdot\arcs4$ &    39 & 8.2 &  0.1 \\
LMC0211 & $5\uph12\upm03\zdot\ups79$ & $-69\arcd12\arcm53\zdot\arcs5$ &    26 & 8.2 &  0.05 \\
\hline\xrule
LMC0210 & $5\uph12\upm03\zdot\ups06$ & $-69\arcd17\arcm11\zdot\arcs8$ &    18 & 7.9 &  0.1 \\
LMC0212 & $5\uph12\upm08\zdot\ups79$ & $-69\arcd16\arcm44\zdot\arcs5$ &    20 & 7.9 &  0.1 \\
LMC0218 & $5\uph12\upm17\zdot\ups18$ & $-69\arcd17\arcm31\zdot\arcs9$ &   13 & 9.1 &  0.1 \\ 
LMC0219 & $5\uph12\upm18\zdot\ups11$ & $-69\arcd17\arcm02\zdot\arcs9$ &   19 & 8.0 &  0.05 \\ 
\hline\xrule
LMC0213 & $5\uph12\upm09\zdot\ups46$ & $-68\arcd54\arcm44\zdot\arcs3$ &    12 & 8.9 &  0.05 \\
LMC0215 & $5\uph12\upm14\zdot\ups91$ & $-68\arcd55\arcm52\zdot\arcs1$ &    18 & 8.2 &  0.08 \\
LMC0214 & $5\uph12\upm13\zdot\ups20$ & $-68\arcd57\arcm04\zdot\arcs5$ &    18 & 8.4 &  0.1 \\
\hline\xrule
LMC0216 & $5\uph12\upm14\zdot\ups92$ & $-69\arcd25\arcm03\zdot\arcs6$ &    23 & 8.9 &  0.1 \\
LMC0220 & $5\uph12\upm21\zdot\ups16$ & $-69\arcd24\arcm41\zdot\arcs3$ &    14 & 8.05 &  0.05 \\
\hline\xrule
LMC0226 & $5\uph12\upm34\zdot\ups43$ & $-69\arcd17\arcm13\zdot\arcs7$ &    16 & 7.6 &  0.08 \\
LMC0227 & $5\uph12\upm38\zdot\ups05$ & $-69\arcd17\arcm33\zdot\arcs0$ &    14 & 7.9 &  0.1 \\
\hline\xrule
LMC0232 & $5\uph12\upm57\zdot\ups60$ & $-69\arcd04\arcm05\zdot\arcs7$ &    12 & 8.25 &  0.05 \\
LMC0233 & $5\uph13\upm03\zdot\ups60$ & $-69\arcd02\arcm59\zdot\arcs6$ &    16 & 9.05 &  0.05 \\
\hline\xrule
LMC0237 & $5\uph13\upm13\zdot\ups22$ & $-69\arcd22\arcm30\zdot\arcs3$ &    18 & $<6.7$ &  -- \\
LMC0238 & $5\uph13\upm19\zdot\ups04$ & $-69\arcd21\arcm44\zdot\arcs5$ &    32 & $<6.7$ &  -- \\
LMC0240 & $5\uph13\upm21\zdot\ups75$ & $-69\arcd22\arcm37\zdot\arcs9$ &    27 & 7.0 &  0.1 \\
LMC0242 & $5\uph13\upm28\zdot\ups42$ & $-69\arcd22\arcm21\zdot\arcs7$ &    23 & $<6.7$ &  -- \\
LMC0246 & $5\uph13\upm38\zdot\ups90$ & $-69\arcd23\arcm02\zdot\arcs0$ &    18 & $<6.8$ &  -- \\
LMC0247 & $5\uph13\upm40\zdot\ups08$ & $-69\arcd22\arcm26\zdot\arcs8$ &    11 & 7.8 &  0.1 \\
\hline\xrule
LMC0266 & $5\uph15\upm27\zdot\ups32$ & $-69\arcd20\arcm43\zdot\arcs0$ &    18 & 8.0 &  0.1 \\
LMC0274 & $5\uph15\upm40\zdot\ups46$ & $-69\arcd20\arcm18\zdot\arcs2$ &    13 & 7.95 &  0.05 \\
\hline\xrule
LMC0267 & $5\uph15\upm33\zdot\ups35$ & $-69\arcd31\arcm56\zdot\arcs5$ &     9 & -- &  -- \\
LMC0275 & $5\uph15\upm44\zdot\ups20$ & $-69\arcd32\arcm25\zdot\arcs8$ &    14 & -- &  -- \\
LMC0279 & $5\uph15\upm54\zdot\ups81$ & $-69\arcd32\arcm14\zdot\arcs2$ &    13 & -- &  -- \\
\hline\xrule
LMC0270 & $5\uph15\upm37\zdot\ups18$ & $-69\arcd28\arcm24\zdot\arcs5$ &    25 & 6.7 &  0.1 \\
LMC0278 & $5\uph15\upm52\zdot\ups01$ & $-69\arcd28\arcm08\zdot\arcs2$ &    31 & 7.85 &  0.05 \\
LMC0280 & $5\uph15\upm56\zdot\ups96$ & $-69\arcd27\arcm16\zdot\arcs4$ &    19 & -- &  -- \\
\hline\xrule
LMC0294 & $5\uph16\upm48\zdot\ups94$ & $-69\arcd34\arcm50\zdot\arcs0$ &    16 & -- &  -- \\
LMC0301 & $5\uph16\upm54\zdot\ups05$ & $-69\arcd34\arcm56\zdot\arcs3$ &    14 & 8.2 &  0.1 \\
\hline\xrule
LMC0296 & $5\uph16\upm50\zdot\ups15$ & $-69\arcd03\arcm35\zdot\arcs0$ &    11 & -- &  -- \\
LMC0297 & $5\uph16\upm52\zdot\ups07$ & $-69\arcd04\arcm13\zdot\arcs4$ &     7 & -- &  -- \\
\hline\xrule
LMC0298 & $5\uph16\upm52\zdot\ups88$ & $-69\arcd09\arcm00\zdot\arcs0$ &    12 & 7.3 &  0.2 \\
LMC0303 & $5\uph16\upm55\zdot\ups59$ & $-69\arcd08\arcm51\zdot\arcs2$ &    30 & 6.7 &  0.1 \\
\hline}
\setcounter{table}{0}
\MakeTableSep{|c|c|c|c|c|c|}{12.5cm}{continued}
{\hline\xrule
Name  &  $\alpha_{2000}$ & $\delta_{2000}$ & Radius
& $\log t$ & $\sigma_{\log t}$\\
OGLE-CL-& & &[\arcs]&\\
\hline\xrule
LMC0302 & $5\uph16\upm54\zdot\ups41$ & $-68\arcd52\arcm35\zdot\arcs8$ &    20 & 8.65 &  0.05 \\
LMC0304 & $5\uph17\upm08\zdot\ups00$ & $-68\arcd52\arcm23\zdot\arcs5$ &    41 & 9.0 &  0.1 \\
\hline\xrule
LMC0309 & $5\uph17\upm22\zdot\ups39$ & $-69\arcd20\arcm16\zdot\arcs2$ &    49 & 7.88 &  0.03 \\
LMC0311 & $5\uph17\upm26\zdot\ups59$ & $-69\arcd22\arcm31\zdot\arcs8$ &    45 & 9.0 &  0.08 \\
LMC0312 & $5\uph17\upm27\zdot\ups68$ & $-69\arcd21\arcm22\zdot\arcs3$ &    31 & 8.0 &  0.05 \\
\hline\xrule
LMC0316 & $5\uph17\upm43\zdot\ups83$ & $-69\arcd34\arcm06\zdot\arcs1$ &    22 & 6.8 &  0.2 \\
LMC0317 & $5\uph17\upm45\zdot\ups94$ & $-69\arcd34\arcm24\zdot\arcs4$ &    16 & -- &  -- \\
LMC0321 & $5\uph17\upm56\zdot\ups16$ & $-69\arcd34\arcm52\zdot\arcs3$ &    17 & 8.1 &  0.07 \\
\hline\xrule
LMC0324 & $5\uph18\upm06\zdot\ups44$ & $-69\arcd31\arcm46\zdot\arcs4$ &    29 & 7.4 &  0.1 \\
LMC0326 & $5\uph18\upm10\zdot\ups51$ & $-69\arcd32\arcm26\zdot\arcs8$ &    14 & 8.0 &  0.1 \\
LMC0330 & $5\uph18\upm18\zdot\ups74$ & $-69\arcd32\arcm14\zdot\arcs8$ &    41 & 7.38 &  0.03 \\
\hline\xrule
LMC0329 & $5\uph18\upm18\zdot\ups05$ & $-69\arcd45\arcm04\zdot\arcs9$ &    18 & 8.0 &  0.1 \\
LMC0334 & $5\uph18\upm31\zdot\ups19$ & $-69\arcd45\arcm14\zdot\arcs6$ &    24 & 7.95 &  0.08 \\
\hline\xrule
LMC0338 & $5\uph18\upm42\zdot\ups53$ & $-69\arcd14\arcm12\zdot\arcs3$ &    46 & 6.7 &  0.1 \\
LMC0340 & $5\uph18\upm46\zdot\ups72$ & $-69\arcd13\arcm32\zdot\arcs4$ &    19 & $<6.8$ &  -- \\
\hline\xrule
LMC0351 & $5\uph19\upm25\zdot\ups74$ & $-69\arcd32\arcm27\zdot\arcs1$ &    33 & 8.3 &  0.05 \\
LMC0353 & $5\uph19\upm33\zdot\ups88$ & $-69\arcd32\arcm31\zdot\arcs9$ &    26 & 8.8 &  0.1 \\
\hline\xrule
LMC0359 & $5\uph19\upm57\zdot\ups48$ & $-69\arcd25\arcm02\zdot\arcs8$ &    18 & 8.0 &  0.1 \\
LMC0362 & $5\uph20\upm03\zdot\ups01$ & $-69\arcd23\arcm59\zdot\arcs1$ &    11 & 9 &  0.05 \\
\hline\xrule
LMC0361 & $5\uph20\upm02\zdot\ups05$ & $-69\arcd15\arcm39\zdot\arcs6$ &    10 & 8.0 &  0.1 \\
LMC0363 & $5\uph20\upm04\zdot\ups43$ & $-69\arcd15\arcm54\zdot\arcs6$ &     9 & -- &  -- \\
\hline\xrule
LMC0365 & $5\uph20\upm08\zdot\ups04$ & $-70\arcd09\arcm15\zdot\arcs0$ &    10 & 8.3 &  0.1 \\
LMC0366 & $5\uph20\upm08\zdot\ups08$ & $-70\arcd08\arcm34\zdot\arcs0$ &    10 & -- &  -- \\
\hline\xrule
LMC0367 & $5\uph20\upm15\zdot\ups93$ & $-69\arcd20\arcm24\zdot\arcs8$ &    14 & 8.0 &  0.1 \\
LMC0370 & $5\uph20\upm25\zdot\ups45$ & $-69\arcd21\arcm18\zdot\arcs1$ &    20 & 7.8 &  0.1 \\
LMC0372 & $5\uph20\upm27\zdot\ups62$ & $-69\arcd21\arcm53\zdot\arcs3$ &    20 & 8.55 &  0.1 \\
\hline\xrule
LMC0369 & $5\uph20\upm23\zdot\ups57$ & $-69\arcd35\arcm03\zdot\arcs1$ &    28 & 8.3 &  0.07 \\
LMC0371 & $5\uph20\upm25\zdot\ups83$ & $-69\arcd34\arcm12\zdot\arcs7$ &    26 & -- &  -- \\
\hline\xrule
LMC0375 & $5\uph20\upm30\zdot\ups61$ & $-69\arcd32\arcm09\zdot\arcs0$ &    31 & 7.7 &  0.05 \\
LMC0379 & $5\uph20\upm35\zdot\ups42$ & $-69\arcd31\arcm32\zdot\arcs9$ &    27 & 8.0 &  0.1 \\
\hline\xrule
LMC0382 & $5\uph20\upm57\zdot\ups73$ & $-69\arcd28\arcm40\zdot\arcs2$ &    31 & 9.03 &  0.05 \\
LMC0383 & $5\uph20\upm59\zdot\ups72$ & $-69\arcd29\arcm44\zdot\arcs8$ &    11 & -- &  -- \\
\hline\xrule
LMC0389 & $5\uph21\upm10\zdot\ups93$ & $-69\arcd56\arcm36\zdot\arcs8$ &    16 & 8.7 &  0.08 \\
LMC0394 & $5\uph21\upm24\zdot\ups45$ & $-69\arcd56\arcm27\zdot\arcs5$ &    25 & 8.55 &  0.05 \\
LMC0395 & $5\uph21\upm26\zdot\ups82$ & $-69\arcd56\arcm59\zdot\arcs0$ &    30 & 9.0 &  0.1 \\
\hline\xrule
LMC0390 & $5\uph21\upm18\zdot\ups65$ & $-69\arcd28\arcm35\zdot\arcs7$ &    17 & 9.0 &  0.08 \\
LMC0393 & $5\uph21\upm23\zdot\ups54$ & $-69\arcd29\arcm26\zdot\arcs6$ &    18 & -- &  -- \\
\hline\xrule
LMC0404 & $5\uph22\upm03\zdot\ups30$ & $-69\arcd15\arcm17\zdot\arcs9$ &    24 & 8.35 &  0.05 \\
LMC0405 & $5\uph22\upm06\zdot\ups85$ & $-69\arcd14\arcm44\zdot\arcs7$ &    14 & 8.3 &  0.1 \\
\hline\xrule
LMC0407 & $5\uph22\upm14\zdot\ups67$ & $-69\arcd30\arcm40\zdot\arcs7$ &    40 & 8.2 &  0.1 \\
LMC0408 & $5\uph22\upm26\zdot\ups26$ & $-69\arcd29\arcm53\zdot\arcs5$ &    35 & -- &  -- \\
\hline}
\setcounter{table}{0}
\MakeTableSep{|c|c|c|c|c|c|}{12.5cm}{continued}
{\hline\xrule
Name  &  $\alpha_{2000}$ & $\delta_{2000}$ & Radius
& $\log t$ & $\sigma_{\log t}$\\
OGLE-CL-& & &[\arcs]&\\
\hline\xrule
LMC0409 & $5\uph22\upm27\zdot\ups28$ & $-69\arcd44\arcm43\zdot\arcs0$ &    24 & 7.55 &  0.05 \\
LMC0413 & $5\uph22\upm37\zdot\ups90$ & $-69\arcd44\arcm39\zdot\arcs9$ &    22 & 8.75 &  0.05 \\
\hline\xrule
LMC0417 & $5\uph23\upm12\zdot\ups94$ & $-69\arcd49\arcm23\zdot\arcs0$ &    22 & 8.3 &  0.1 \\
LMC0418 & $5\uph23\upm19\zdot\ups32$ & $-69\arcd49\arcm46\zdot\arcs5$ &    16 & -- &  -- \\
LMC0419 & $5\uph23\upm25\zdot\ups24$ & $-69\arcd50\arcm07\zdot\arcs1$ &    26 & 7.98 &  0.05 \\
LMC0425 & $5\uph23\upm36\zdot\ups75$ & $-69\arcd49\arcm18\zdot\arcs7$ &    30 & -- &  -- \\
\hline\xrule
LMC0422 & $5\uph23\upm32\zdot\ups17$ & $-69\arcd54\arcm14\zdot\arcs0$ &    11 & 8.2 &  0.1 \\
LMC0424 & $5\uph23\upm35\zdot\ups48$ & $-69\arcd54\arcm17\zdot\arcs7$ &    10 & 8.0 &  0.1 \\
\hline\xrule
LMC0430 & $5\uph24\upm16\zdot\ups37$ & $-69\arcd39\arcm12\zdot\arcs9$ &    12 & 8.8 &  0.05 \\
LMC0433 & $5\uph24\upm21\zdot\ups58$ & $-69\arcd38\arcm28\zdot\arcs9$ &    11 & 9.0 &  0.1 \\
\hline\xrule
LMC0431 & $5\uph24\upm20\zdot\ups42$ & $-69\arcd46\arcm26\zdot\arcs4$ &    19 & 8.0 &  0.05 \\
LMC0434 & $5\uph24\upm23\zdot\ups94$ & $-69\arcd46\arcm47\zdot\arcs5$ &    10 & 9.0 &  0.1 \\
\hline\xrule
LMC0436 & $5\uph24\upm33\zdot\ups04$ & $-69\arcd54\arcm04\zdot\arcs3$ &    44 & 8.7 &  0.08 \\
LMC0437 & $5\uph24\upm33\zdot\ups45$ & $-69\arcd55\arcm26\zdot\arcs9$ &    14 & 8.7 &  0.1 \\
LMC0440 & $5\uph24\upm41\zdot\ups59$ & $-69\arcd53\arcm10\zdot\arcs8$ &    21 & 8.1 &  0.08 \\
\hline\xrule
LMC0441 & $5\uph24\upm52\zdot\ups46$ & $-69\arcd50\arcm36\zdot\arcs7$ &    11 & -- &  -- \\
LMC0442 & $5\uph24\upm53\zdot\ups02$ & $-69\arcd49\arcm47\zdot\arcs2$ &    29 & 8.1 &  0.1 \\
LMC0443 & $5\uph24\upm55\zdot\ups33$ & $-69\arcd50\arcm13\zdot\arcs9$ &    22 & 8.0 &  0.03 \\
LMC0444 & $5\uph24\upm55\zdot\ups46$ & $-69\arcd51\arcm46\zdot\arcs0$ &    14 & 8.0 &  0.08 \\
LMC0447 & $5\uph25\upm03\zdot\ups54$ & $-69\arcd52\arcm12\zdot\arcs7$ &    12 & 8.1 &  0.1 \\
LMC0449 & $5\uph25\upm05\zdot\ups90$ & $-69\arcd52\arcm27\zdot\arcs2$ &    36 & -- &  -- \\
\hline\xrule
LMC0445 & $5\uph24\upm56\zdot\ups68$ & $-69\arcd25\arcm29\zdot\arcs3$ &    18 & 8.6 &  0.1 \\
LMC0446 & $5\uph25\upm01\zdot\ups13$ & $-69\arcd26\arcm03\zdot\arcs1$ &    35 & 8.3 &  0.05 \\
\hline\xrule
LMC0448 & $5\uph25\upm04\zdot\ups69$ & $-69\arcd44\arcm14\zdot\arcs3$ &    20 & 8.6 &  0.1 \\
LMC0450 & $5\uph25\upm06\zdot\ups87$ & $-69\arcd42\arcm56\zdot\arcs3$ &    20 & 8.2 &  0.1 \\
\hline\xrule
LMC0454 & $5\uph25\upm23\zdot\ups00$ & $-69\arcd47\arcm07\zdot\arcs0$ &    16 & 8.6,9.0 &  0.05 \\
LMC0456 & $5\uph25\upm28\zdot\ups00$ & $-69\arcd46\arcm31\zdot\arcs6$ &    21 & 8.45 &  0.05 \\
\hline\xrule
LMC0457 & $5\uph25\upm30\zdot\ups72$ & $-69\arcd50\arcm09\zdot\arcs6$ &    37 & 8.0 &  0.1 \\
LMC0461 & $5\uph25\upm38\zdot\ups49$ & $-69\arcd49\arcm30\zdot\arcs8$ &    30 & 7.95 &  0.05 \\
\hline\xrule
LMC0465 & $5\uph25\upm53\zdot\ups77$ & $-69\arcd46\arcm13\zdot\arcs5$ &    20 & 8.8 &  0.05 \\
LMC0467 & $5\uph25\upm57\zdot\ups30$ & $-69\arcd45\arcm03\zdot\arcs9$ &    22 & 8.3 &  0.05 \\
\hline\xrule
LMC0475 & $5\uph26\upm30\zdot\ups13$ & $-69\arcd47\arcm26\zdot\arcs0$ &    12 & -- &  -- \\
LMC0476 & $5\uph26\upm33\zdot\ups08$ & $-69\arcd48\arcm12\zdot\arcs0$ &    23 & 8.4 &  0.1 \\
LMC0477 & $5\uph26\upm34\zdot\ups11$ & $-69\arcd50\arcm26\zdot\arcs7$ &    37 & 7.8 &  0.05 \\
LMC0478 & $5\uph26\upm35\zdot\ups30$ & $-69\arcd49\arcm23\zdot\arcs1$ &    30 & 8.0 &  0.1 \\
LMC0480 & $5\uph26\upm45\zdot\ups58$ & $-69\arcd51\arcm03\zdot\arcs2$ &    31 & 8.0 &  0.05 \\
LMC0481 & $5\uph26\upm48\zdot\ups80$ & $-69\arcd50\arcm17\zdot\arcs2$ &    29 & 7.8 &  0.1 \\
\hline\xrule
LMC0482 & $5\uph26\upm52\zdot\ups66$ & $-69\arcd46\arcm03\zdot\arcs0$ &    22 & 8.9 &  9.0 \\
LMC0485 & $5\uph27\upm00\zdot\ups68$ & $-69\arcd46\arcm37\zdot\arcs5$ &    25 & 7.95 &  0.05 \\
\hline\xrule
LMC0495 & $5\uph27\upm35\zdot\ups63$ & $-69\arcd53\arcm49\zdot\arcs6$ &    19 & 8.6 &  0.2 \\
LMC0497 & $5\uph27\upm47\zdot\ups58$ & $-69\arcd53\arcm29\zdot\arcs8$ &    24 & 8.1 &  0.05 \\
\hline}
\setcounter{table}{0}
\MakeTableSep{|c|c|c|c|c|c|}{12.5cm}{continued}
{\hline\xrule
Name  &  $\alpha_{2000}$ & $\delta_{2000}$ & Radius
& $\log t$ & $\sigma_{\log t}$\\
OGLE-CL-& & &[\arcs]&\\
\hline\xrule
LMC0500 & $5\uph28\upm05\zdot\ups10$ & $-69\arcd59\arcm16\zdot\arcs7$ &    12 & 8.2 &  0.1 \\
LMC0501 & $5\uph28\upm06\zdot\ups85$ & $-70\arcd00\arcm08\zdot\arcs7$ &    12 & -- &  -- \\
\hline\xrule
LMC0504 & $5\uph28\upm25\zdot\ups20$ & $-69\arcd57\arcm12\zdot\arcs0$ &    25 & 7.65 &  0.05 \\
LMC0510 & $5\uph28\upm41\zdot\ups10$ & $-69\arcd57\arcm13\zdot\arcs0$ &    20 & 8.0 &  0.05 \\
\hline\xrule
LMC0505 & $5\uph28\upm26\zdot\ups78$ & $-69\arcd46\arcm05\zdot\arcs3$ &    15 & 8.4 &  0.05 \\
LMC0511 & $5\uph28\upm42\zdot\ups33$ & $-69\arcd46\arcm06\zdot\arcs4$ &    31 & 9.0 &  0.1 \\
\hline\xrule
LMC0507 & $5\uph28\upm31\zdot\ups72$ & $-69\arcd50\arcm32\zdot\arcs1$ &    16 & 8.7 &  0.1 \\
LMC0509 & $5\uph28\upm40\zdot\ups97$ & $-69\arcd49\arcm51\zdot\arcs0$ &    12 & -- &  -- \\
LMC0512 & $5\uph28\upm44\zdot\ups44$ & $-69\arcd50\arcm04\zdot\arcs9$ &    19 & 8.1 &  0.05 \\
\hline\xrule
LMC0517 & $5\uph29\upm18\zdot\ups80$ & $-69\arcd54\arcm52\zdot\arcs5$ &    16 & 8.2 &  0.15 \\
LMC0520 & $5\uph29\upm24\zdot\ups59$ & $-69\arcd55\arcm11\zdot\arcs8$ &    18 & 8.2 &  0.05 \\
\hline\xrule
LMC0521 & $5\uph29\upm27\zdot\ups00$ & $-69\arcd47\arcm06\zdot\arcs4$ &    11 & -- &  -- \\
LMC0525 & $5\uph29\upm34\zdot\ups59$ & $-69\arcd46\arcm32\zdot\arcs8$ &    20 & 8.1 &  0.1 \\
\hline\xrule
LMC0529 & $5\uph29\upm59\zdot\ups77$ & $-70\arcd03\arcm41\zdot\arcs6$ &    12 & -- &  -- \\
LMC0539 & $5\uph30\upm11\zdot\ups38$ & $-70\arcd04\arcm09\zdot\arcs7$ &    14 & -- &  -- \\
\hline\xrule
LMC0530 & $5\uph29\upm59\zdot\ups95$ & $-69\arcd31\arcm21\zdot\arcs3$ &    12 & 9.0 &  0.1 \\
LMC0531 & $5\uph30\upm00\zdot\ups73$ & $-69\arcd31\arcm37\zdot\arcs1$ &    14 & 9.1 &  0.05 \\
\hline\xrule
LMC0532 & $5\uph30\upm01\zdot\ups73$ & $-69\arcd57\arcm02\zdot\arcs3$ &    16 & 8.3 &  0.05 \\
LMC0533 & $5\uph30\upm01\zdot\ups93$ & $-69\arcd56\arcm38\zdot\arcs2$ &    16 & -- &  -- \\
\hline\xrule
LMC0537 & $5\uph30\upm04\zdot\ups37$ & $-69\arcd44\arcm27\zdot\arcs4$ &    11 & -- &  -- \\
LMC0538 & $5\uph30\upm10\zdot\ups37$ & $-69\arcd45\arcm09\zdot\arcs6$ &    57 & $>9.2$ &  -- \\
\hline\xrule
LMC0542 & $5\uph30\upm34\zdot\ups20$ & $-70\arcd11\arcm51\zdot\arcs4$ &    20 & -- &  -- \\
LMC0545 & $5\uph30\upm39\zdot\ups55$ & $-70\arcd13\arcm06\zdot\arcs9$ &    11 & 8.9 &  0.1 \\
LMC0546 & $5\uph30\upm40\zdot\ups70$ & $-70\arcd13\arcm21\zdot\arcs2$ &    14 & 8.9 &  0.05 \\
\hline\xrule
LMC0555 & $5\uph31\upm19\zdot\ups49$ & $-70\arcd01\arcm59\zdot\arcs6$ &    13 & -- &  -- \\
LMC0558 & $5\uph31\upm30\zdot\ups77$ & $-70\arcd01\arcm24\zdot\arcs5$ &    18 & 8.8 &  0.1 \\
\hline\xrule
LMC0562 & $5\uph31\upm45\zdot\ups78$ & $-70\arcd18\arcm27\zdot\arcs0$ &    11 & -- &  -- \\
LMC0564 & $5\uph31\upm50\zdot\ups36$ & $-70\arcd17\arcm21\zdot\arcs5$ &    20 & 8.4 &  0.1 \\
\hline\xrule
LMC0565 & $5\uph31\upm56\zdot\ups48$ & $-70\arcd09\arcm32\zdot\arcs5$ &    49 & $>9.2$ &  -- \\
LMC0566 & $5\uph32\upm01\zdot\ups06$ & $-70\arcd10\arcm42\zdot\arcs6$ &    21 & 8.2 &  0.1 \\
\hline\xrule
LMC0572 & $5\uph32\upm42\zdot\ups62$ & $-69\arcd53\arcm10\zdot\arcs8$ &    27 & -- &  -- \\
LMC0574 & $5\uph32\upm46\zdot\ups01$ & $-69\arcd52\arcm04\zdot\arcs6$ &    14 & 8.3 &  0.1 \\
\hline\xrule
LMC0573 & $5\uph32\upm45\zdot\ups92$ & $-70\arcd26\arcm03\zdot\arcs4$ &    11 & -- &  -- \\
LMC0576 & $5\uph32\upm48\zdot\ups76$ & $-70\arcd26\arcm07\zdot\arcs4$ &    10 & $>8.9$ &  -- \\
LMC0577 & $5\uph32\upm48\zdot\ups86$ & $-70\arcd27\arcm23\zdot\arcs0$ &    25 & 8.2 &  0.05 \\
LMC0578 & $5\uph32\upm51\zdot\ups25$ & $-70\arcd26\arcm01\zdot\arcs5$ &    14 & 8.65 &  0.05 \\
\hline\xrule
LMC0598 & $5\uph34\upm00\zdot\ups48$ & $-69\arcd40\arcm21\zdot\arcs8$ &    30 & 8.5 &  0.1 \\
LMC0601 & $5\uph34\upm14\zdot\ups51$ & $-69\arcd40\arcm34\zdot\arcs1$ &     8 & -- &  -- \\
\hline\xrule
LMC0605 & $5\uph34\upm40\zdot\ups36$ & $-69\arcd44\arcm50\zdot\arcs1$ &    24 & 6.7 &  0.05 \\
LMC0608 & $5\uph34\upm46\zdot\ups65$ & $-69\arcd44\arcm35\zdot\arcs2$ &    23 & 8.05 &  0.05 \\
\hline}
\setcounter{table}{0}
\MakeTableSep{|c|c|c|c|c|c|}{12.5cm}{continued}
{\hline\xrule
Name  &  $\alpha_{2000}$ & $\delta_{2000}$ & Radius
& $\log t$ & $\sigma_{\log t}$\\
OGLE-CL-& & &[\arcs]&\\
\hline\xrule
LMC0616 & $5\uph35\upm14\zdot\ups02$ & $-69\arcd54\arcm21\zdot\arcs2$ &    24 & 6.9 &  0.2 \\
LMC0617 & $5\uph35\upm17\zdot\ups10$ & $-69\arcd54\arcm50\zdot\arcs3$ &    20 & -- &  -- \\
\hline\xrule
LMC0619 & $5\uph35\upm30\zdot\ups68$ & $-70\arcd20\arcm56\zdot\arcs9$ &    14 & -- &  -- \\
LMC0621 & $5\uph35\upm36\zdot\ups65$ & $-70\arcd22\arcm11\zdot\arcs2$ &    14 & -- &  -- \\
\hline\xrule
LMC0622 & $5\uph35\upm38\zdot\ups66$ & $-70\arcd14\arcm23\zdot\arcs1$ &    30 & -- &  -- \\
LMC0625 & $5\uph35\upm51\zdot\ups58$ & $-70\arcd13\arcm51\zdot\arcs2$ &    10 & -- &  -- \\
\hline\xrule
LMC0638 & $5\uph37\upm15\zdot\ups39$ & $-69\arcd53\arcm44\zdot\arcs7$ &    18 & -- &  -- \\
LMC0640 & $5\uph37\upm21\zdot\ups73$ & $-69\arcd53\arcm40\zdot\arcs5$ &    13 & -- &  -- \\
\hline\xrule
LMC0641 & $5\uph37\upm22\zdot\ups08$ & $-69\arcd58\arcm21\zdot\arcs2$ &    34 & -- &  -- \\
LMC0642 & $5\uph37\upm22\zdot\ups24$ & $-69\arcd58\arcm56\zdot\arcs0$ &    10 & -- &  -- \\
\hline\xrule
LMC0644 & $5\uph37\upm25\zdot\ups84$ & $-70\arcd13\arcm28\zdot\arcs6$ &    16 & -- &  -- \\
LMC0648 & $5\uph37\upm37\zdot\ups81$ & $-70\arcd13\arcm56\zdot\arcs4$ &    59 & -- &  -- \\
\hline\xrule
LMC0647 & $5\uph37\upm37\zdot\ups00$ & $-70\arcd07\arcm33\zdot\arcs5$ &    24 & -- &  -- \\
LMC0650 & $5\uph37\upm39\zdot\ups15$ & $-70\arcd08\arcm43\zdot\arcs9$ &    37 & -- &  -- \\
LMC0651 & $5\uph37\upm42\zdot\ups36$ & $-70\arcd09\arcm54\zdot\arcs0$ &    27 & 8.1 &  0.05 \\
\hline\xrule
LMC0663 & $5\uph38\upm57\zdot\ups52$ & $-69\arcd59\arcm31\zdot\arcs5$ &    14 & -- &  -- \\
LMC0664 & $5\uph39\upm00\zdot\ups27$ & $-69\arcd59\arcm19\zdot\arcs5$ &    22 & 8.35 &  0.08 \\
\hline\xrule
LMC0665 & $5\uph39\upm05\zdot\ups63$ & $-70\arcd13\arcm46\zdot\arcs9$ &    18 & 8.85 &  0.1 \\
LMC0666 & $5\uph39\upm17\zdot\ups87$ & $-70\arcd13\arcm11\zdot\arcs9$ &    20 & -- &  -- \\
LMC0667 & $5\uph39\upm27\zdot\ups91$ & $-70\arcd12\arcm35\zdot\arcs6$ &     9 & -- &  -- \\
\hline\xrule
LMC0669 & $5\uph39\upm32\zdot\ups97$ & $-69\arcd53\arcm31\zdot\arcs1$ &    10 & -- &  -- \\
LMC0670 & $5\uph39\upm36\zdot\ups01$ & $-69\arcd54\arcm28\zdot\arcs2$ &    11 & 8.8 &  0.1 \\
\hline\xrule
LMC0679 & $5\uph40\upm56\zdot\ups60$ & $-70\arcd51\arcm27\zdot\arcs7$ &    18 & 9.0 &  0.1 \\
LMC0680 & $5\uph41\upm01\zdot\ups85$ & $-70\arcd50\arcm50\zdot\arcs2$ &    14 & 8.3 &  0.1 \\
\hline\xrule
LMC0686 & $5\uph41\upm29\zdot\ups28$ & $-70\arcd13\arcm58\zdot\arcs0$ &    14 & 8.0 &  0.1 \\
LMC0687 & $5\uph41\upm33\zdot\ups03$ & $-70\arcd14\arcm08\zdot\arcs0$ &    11 & -- &  -- \\
\hline\xrule
LMC0704 & $5\uph43\upm41\zdot\ups52$ & $-70\arcd36\arcm30\zdot\arcs3$ &    23 & 9.0 &  0.05 \\
LMC0708 & $5\uph43\upm55\zdot\ups64$ & $-70\arcd36\arcm37\zdot\arcs6$ &    11 & 9.0 &  0.1 \\
\hline\xrule
LMC0711 & $5\uph44\upm14\zdot\ups10$ & $-70\arcd39\arcm19\zdot\arcs8$ &    16 & 8.3 &  0.1 \\
LMC0712 & $5\uph44\upm14\zdot\ups50$ & $-70\arcd40\arcm09\zdot\arcs5$ &    20 & 8.3 &  0.1 \\
LMC0713 & $5\uph44\upm16\zdot\ups72$ & $-70\arcd59\arcm59\zdot\arcs1$ &    13 & -- &  -- \\
LMC0715 & $5\uph44\upm33\zdot\ups07$ & $-70\arcd59\arcm35\zdot\arcs3$ &    32 & 8.2 &  0.05 \\
LMC0718 & $5\uph44\upm44\zdot\ups66$ & $-71\arcd00\arcm21\zdot\arcs3$ &     8 & 8.4 &  0.1 \\
\hline\xrule
LMC0717 & $5\uph44\upm42\zdot\ups33$ & $-70\arcd25\arcm31\zdot\arcs0$ &    13 & -- &  -- \\
LMC0720 & $5\uph44\upm47\zdot\ups26$ & $-70\arcd24\arcm21\zdot\arcs9$ &    11 & 8.2 &  0.2 \\
\hline\xrule
LMC0731 & $5\uph45\upm46\zdot\ups36$ & $-70\arcd43\arcm09\zdot\arcs0$ &    16 & 8.7 &  0.1 \\
LMC0732 & $5\uph45\upm59\zdot\ups18$ & $-70\arcd43\arcm45\zdot\arcs8$ &     9 & 7.8 &  0.2 \\
LMC0732 & $5\uph45\upm59\zdot\ups18$ & $-70\arcd43\arcm45\zdot\arcs8$ &     9 & 7.8 &  0.2 \\
LMC0733 & $5\uph46\upm11\zdot\ups08$ & $-70\arcd43\arcm12\zdot\arcs2$ &    10 & -- &  -- \\
\hline\xrule
LMC0736 & $5\uph46\upm41\zdot\ups10$ & $-70\arcd50\arcm51\zdot\arcs8$ &    11 & 8.25 &  0.08 \\
LMC0737 & $5\uph46\upm47\zdot\ups18$ & $-70\arcd49\arcm58\zdot\arcs5$ &    15 & 8.35 &  0.05 \\
\hline}

The number of cluster pairs expected from chance line-up due to projection may 
be calculated from the following formula given by Page (1972). 
$$N_1=0.5\pi\times N_2^2\times s^2$$ 
where $N_1$, $N_2$ and $s$ are the expected number of pairs per square degree, 
the number of clusters per square degree and projected angular separation in 
degrees, respectively. Under assumption that 745 clusters from the OGLE 
catalog are distributed uniformly in the 5.8 square degrees region of the LMC 
we find that the number of chance pairs with the separations smaller than 18~pc 
should be 51. Almost the same result, namely 53, was obtained adopting this 
formula to the ${15\times15}$~arcmin regions where the uniform density of 
clusters may be assumed. As we can see, the number of detected pairs, equal to 
153, is significantly larger. Similar results were obtained for the SMC 
multiple clusters (Pietrzy{\'n}ski and Udalski 1999c). 

\Section{Comparison of Multiple Cluster Populations from the LMC and SMC}
\subsection{Multiple Cluster Fraction}
It is very difficult to estimate the completeness of presented catalogs. It 
depends on the completeness of photometric measurements, the completeness of 
catalogs of clusters and the location in a given galaxy. Photometric data 
obtained in the course of the OGLE-II project are complete down to 
${V\approx21.5}$~mag. For that reason we were not able to detect sparely 
populated clusters older than about one billion years. It should be stressed 
that population of potential multiple clusters also depends strongly on the 
adopted distances and separation limits. The OGLE lists of multiple clusters 
in the SMC and the LMC were prepared based on uniform observational data in 
the same, consistent manner so the completeness in both cases should be 
similar. We can see then that the observed fractions of multiple clusters in 
these galaxies (0.11 and 0.13 in the case of the SMC and LMC, respectively) 
are almost the same. 

\subsection{Size Distribution}
Fig.~1 presents the size distribution of multiple cluster components from the 
Magellanic Clouds. Solid and dotted lines represent objects from the LMC, 
described in this paper and from the SMC (Pietrzy{\'n}ski and Udalski 1999c), 
respectively. The shapes of these distributions strongly resemble the shapes 
of the size distribution of single clusters (Pietrzy{\'n}ski 1999). 

Fig.~2 displays distribution of size ratios of the multiple cluster 
components from the Magellanic Clouds. Pronounced peaks located around 1 
indicate that in most cases multiple cluster components have very similar 
sizes. 
\begin{figure}[p]
\includegraphics[width=10cm]{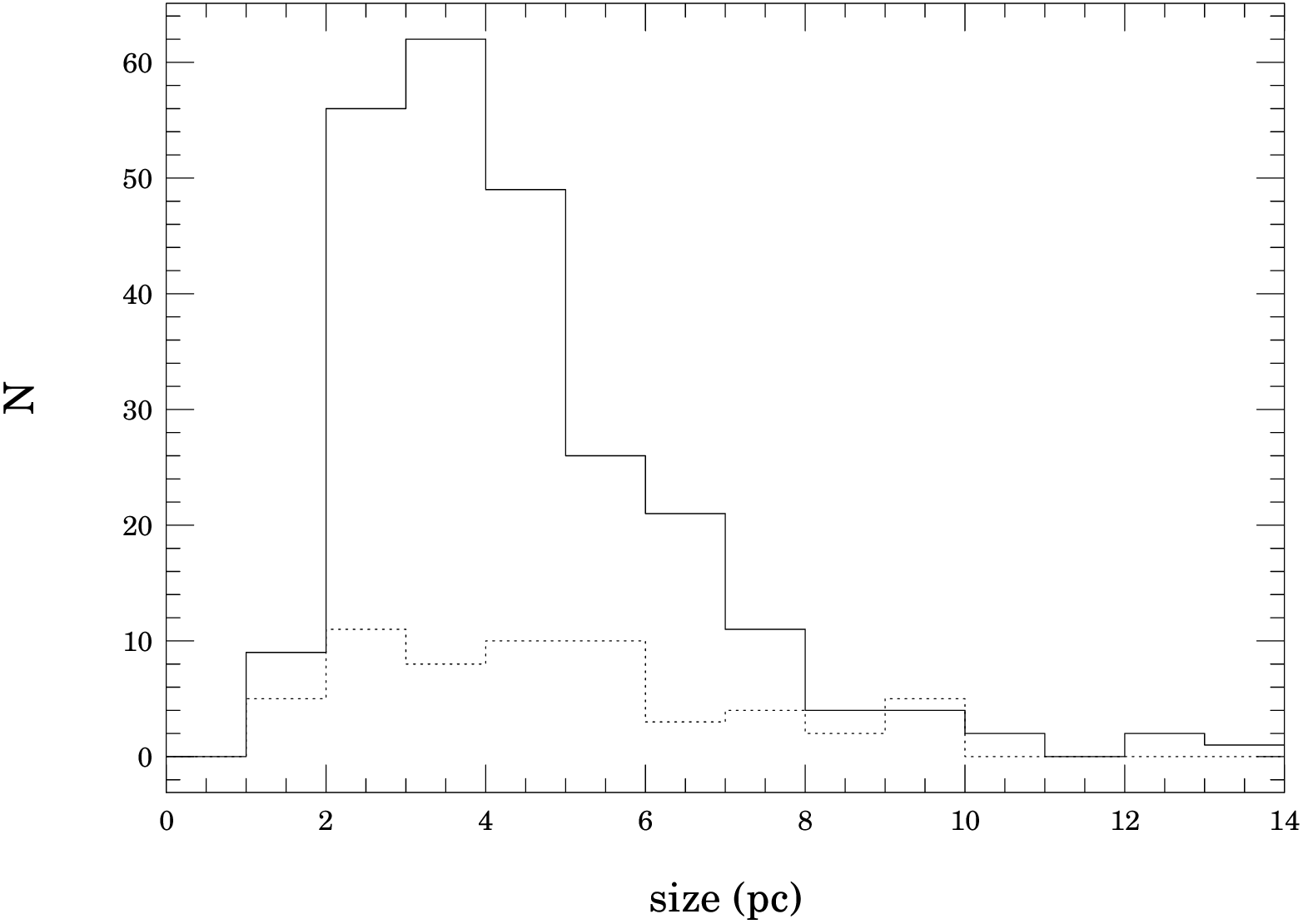}
\FigCap{Size distribution of multiple cluster components from the LMC -- solid 
line, and the SMC -- dotted line.} 
\vskip.3cm
\includegraphics[width=10cm]{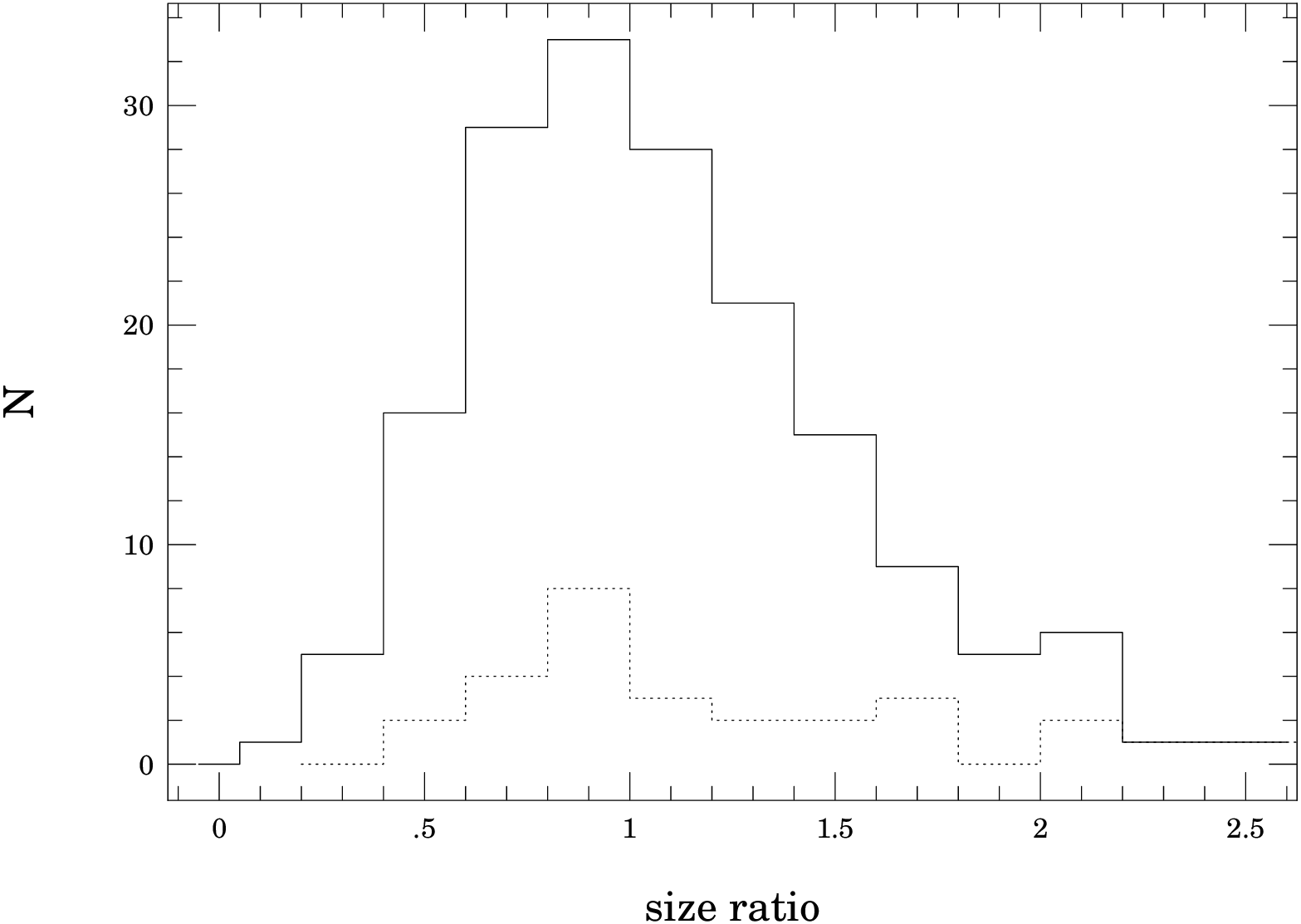}
\FigCap{Distribution of size ratios of multiple cluster components from the 
LMC -- solid line, and the SMC -- dotted line.} 
\end{figure}

\subsection{Distribution of Separations}
Previous study of distribution of separations of 69 cluster pairs from the LMC 
(Bhatia and Hatzidimitriou 1988) revealed two peaks at 5 and 15~pc. In the 
case of the SMC, Hatzidimitriou and Bhatia (1990) detected only one peak 
located around 11~pc. 

Distributions of separations of multiple cluster candidates from the 
Magellanic Clouds obtained with the OGLE data are displayed in Fig.~3. Two 
peaks located at about 9 and 15~pc are visible in both distributions. They are 
also present if one uses different binning. This fact suggests the similar 
mechanisms of multiple clusters formation and disruption in both the SMC and 
LMC. 
\begin{figure}[htb]
\includegraphics[width=10cm]{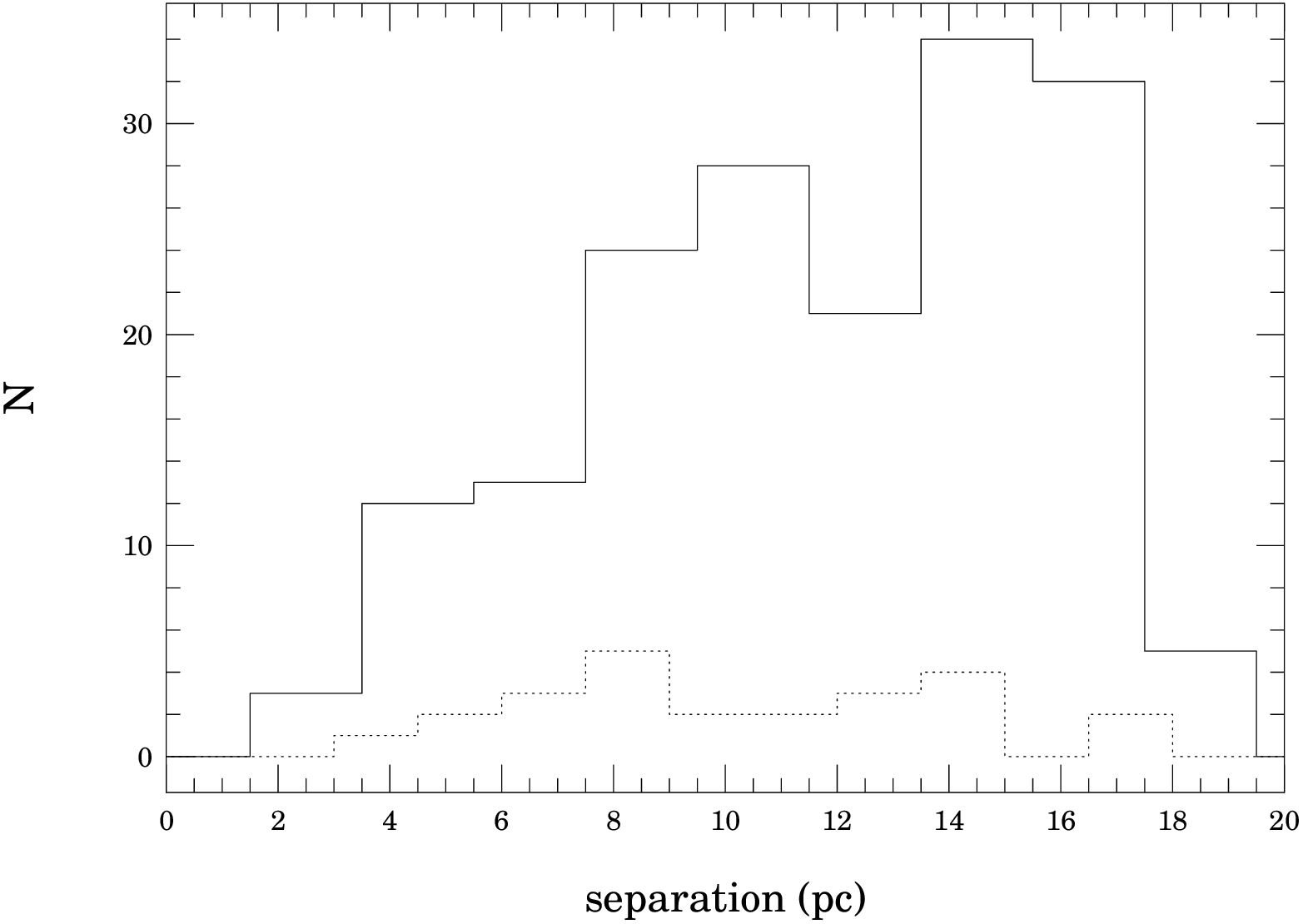}
\FigCap{Distribution of separation of multiple cluster components from the LMC 
-- solid line, and the SMC -- dotted line.} 
\end{figure}

\subsection{Ages of Multiple Clusters}
Ages of presented multiple clusters from the LMC were taken from the list 
of Pietrzy{\'n}ski and Udalski (2000). We found that most, namely 102, 
components of multiple systems have coeval ages. 53 components have 
significantly different ages. In the case of the multiple cluster candidates 
from the SMC, Pietrzy{\'n}ski and Udalski (1999c) found that all of the six 
systems with reliable age determination have coeval ages. Our data indicate 
common origin of most of the components of the multiple systems. Further 
studies are required to check whether the components of multiple systems with 
very different ages are physically bound. 
\begin{figure}[htb]
\vglue-6mm
\includegraphics[width=10cm]{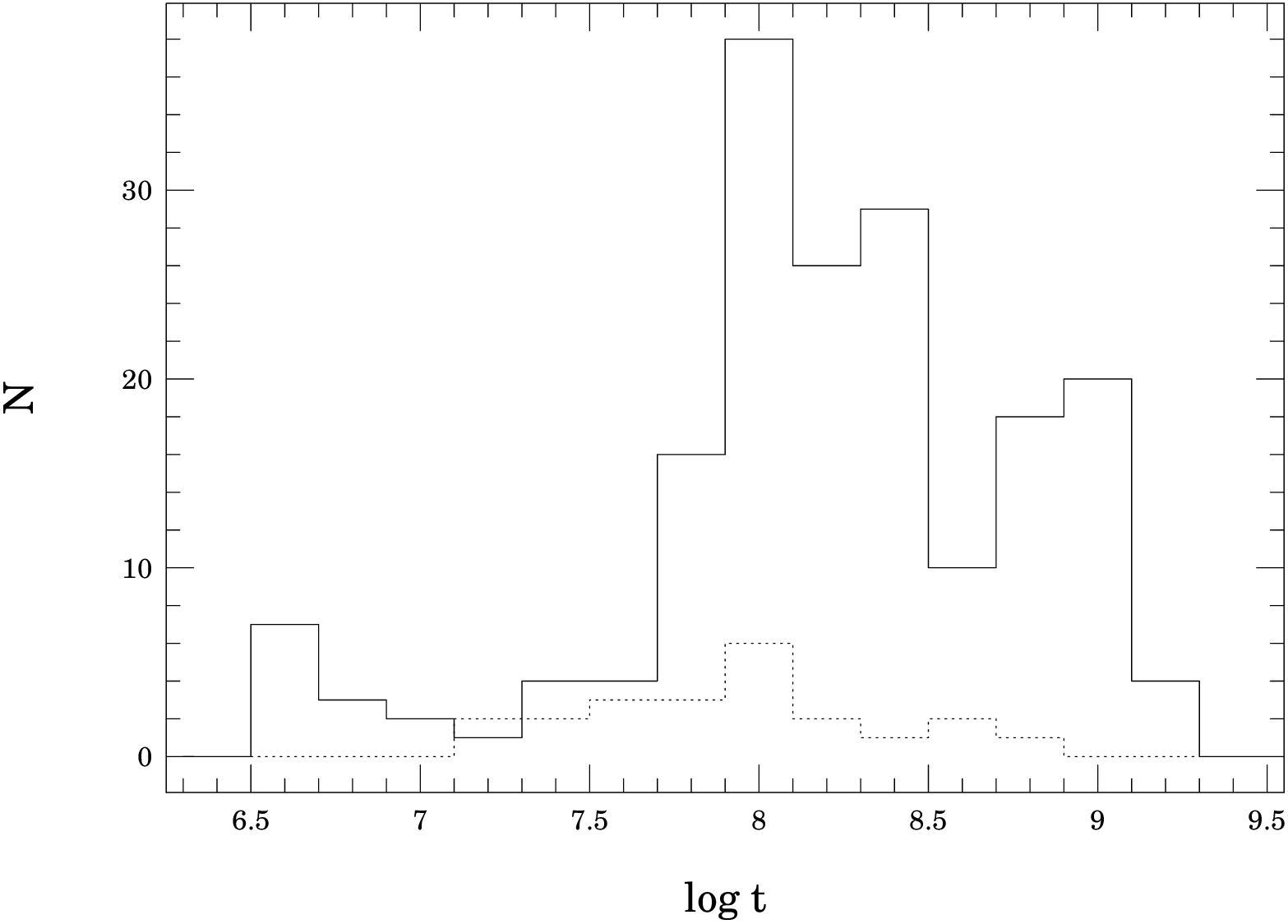}
\FigCap{Distribution of ages of multiple cluster components from the LMC -- 
solid line, and the SMC -- dotted line.} 
\end{figure}

Fig.~4 displays the distribution of ages of potential multiple cluster 
candidate components from the Magellanic Clouds. It is clearly seen that most 
objects are younger than 300 million years. The peaks are seen around 100~Myr 
in both distributions which may be connected with the last encounter between 
the SMC and the LMC (Gardiner \etal 1994). Similar peaks are present in the 
distribution of ages of single clusters from the SMC and the LMC 
(Pietrzy{\'n}ski and Udalski 2000). 

\Section{Summary}
We presented the multiple cluster candidates from the LMC selected from the 
OGLE catalog of star clusters (Pietrzy{\'n}ski and Udalski 1999c). Simple 
statistical considerations show that significant fraction of them may 
constitute physical systems. We compare the populations of potential multiple 
clusters from the LMC and SMC. Taking into account that the completeness of 
these catalogs should be comparable, we found that the fractions of potential 
multiple systems in these galaxies are almost the same, around 12\%. Based on 
inspection of size and age distributions we can conclude that their shapes are 
very similar to the shapes of distributions of ages and sizes of single 
clusters. The difference of sizes between components of a given system is 
small. The distributions of separations reveal two peaks at 9 and 15~pc. Most 
clusters are young. Their distributions of ages show peaks at about 100~Myr. 
Further photometric and spectroscopic studies are necessary to confirm 
physical nature of the presented systems. 

\Acknow{The paper was partly supported by the Polish KBN grants 2P03D00814 to 
A.~Udalski. Partial support for the OGLE project was provided with the NSF 
grant AST-9530478 to B.~Paczy\'nski.}

\end{document}